\RequirePackage{fixltx2e}
\documentclass[aps, prl, superscriptaddress, onecolumn, notitlepage]{revtex4-1}

\usepackage[colorlinks]{hyperref}
\usepackage{lmodern}
\usepackage{amsfonts, amssymb, amsmath}
\usepackage{graphicx}
\usepackage{color}
\usepackage[utf8x]{inputenc}

\newcommand{\uu}{\mathbf{u}}

\newcommand{\uddot}{ü}

\graphicspath{{./figs/}}

\newcommand{\jhuams}{Department of Applied Mathematics \& Statistics, The Johns Hopkins University, Baltimore, MD 21218, USA}
\newcommand{\jhupha}{Department of Physics \& Astronomy, The Johns Hopkins University, Baltimore, MD 21218, USA}
\newcommand{\jhume}{Department of Mechanical Engineering, The Johns Hopkins University, Baltimore, MD 21218, USA}
\newcommand{\tum}{Informatik 15 (Computer Graphik \& Visualisierung), Technische Universit\"at M\uddot nchen, 85748 Garching bei M\uddot nchen, Germany}

\begin{document}
	
	\title{Vortices within vortices: hierarchical nature of vortex tubes in turbulence}

	\author{Kai B\"{u}rger}
	\email{buerger@tum.de}
	\affiliation{\tum}

	\author{Marc Treib}
	\email{treib@tum.de}
	\affiliation{\tum}

	\author{R\"udiger Westermann}
	\email{westermann@tum.de}
	\affiliation{\tum}

	\author{Suzanne Werner}
	\email{swerner@pha.jhu.edu}
	\affiliation{\jhupha}

	\author{Cristian C Lalescu}
	\email{clalesc1@jhu.edu}
%	\thanks{corresponding author}
	\affiliation{\jhuams}

	\author{Alexander Szalay}
	\email{szalay@jhu.edu}
	\affiliation{\jhupha}
	
	\author{Charles Meneveau}
	\email{meneveau@jhu.edu}
	\affiliation{\jhume}
	
	\author{Gregory L Eyink}
	\email{eyink@jhu.edu}
	\affiliation{\jhupha}
	\affiliation{\jhuams}
	\affiliation{\jhume}

	\begin{abstract}
		The JHU turbulence database \cite{JHU_NS_TDB} can be used with a state of the
        art visualisation tool \cite{treib12turbulence} to generate high quality
        \href{https://www.youtube.com/watch?v=hMMsPGihUi0}{fluid dynamics
        videos}.
		In this work we investigate the classical idea that smaller structures in turbulent flows, while engaged in their own internal dynamics, are advected by the larger structures.
		They are not advected undistorted, however.
		We see instead that the small scale structures are sheared and twisted by the larger scales.
		This illuminates the basic mechanisms of the turbulent cascade.
	\end{abstract}

	\maketitle

	\section{The JHU turbulence database}

		In \cite{JHU_NS_TDB} a database containing a solution of the 3D incompressible Navier-Stokes (NS) equations is presented.
		The equations were solved numerically with a standard pseudo-spectral simulation in a periodic domain, using a real space grid of $1024^3$ grid points.
		A large-scale body force drives a turbulent flow with a Taylor microscale based Reynolds number $R_\lambda = 433$.
		Out of this solution, $1024$ snapshots were stored, spread out evenly over a large eddy turnover time.
		More on the simulation and on accessing the data can be found at \url{http://turbulence.pha.jhu.edu}.
		In practical terms, we have easy access to the turbulent velocity field and pressure at every point in space and time.

	\section{Vortices within vortices}

		One usual way of visualising a turbulent velocity field is to plot vorticity isosurfaces --- see for instance the plots from \cite{yokokawa_bigNS_2002}.
		The resulting pictures are usually very ``crowded'', in the sense that there are many intertwined thin vortex tubes, generating an extremely complex structure.
		In fact, the picture of the entire dataset from \cite{yokokawa_bigNS_2002} looks extremely noisy and it is arguably not very informative about the turbulent dynamics.

		In this work, we follow a different approach.
		Firstly, we use the Hunt criterion, i.e. we rely on the alternate quantity
		\begin{equation}
            Q = \frac{1}{4} \left(|\nabla \times \uu|^2 - \tfrac{1}{2}|\nabla \uu + (\nabla \uu)^T|^2\right)
		\end{equation}
        first employed for visualisation in \cite{hunt_eddies_1988}.
		Secondly, the tool being used has the option of displaying data only inside clearly defined domains of 3D space.
		We can exploit this facility to investigate the multiscale character of the turbulent cascade.
		Because vorticity is dominated by the smallest available scales in the velocity, we can visualize vorticity at scale $\ell$ by the curl of the velocity box-filtered at scale $\ell$.
		We follow a simple procedure:
		\begin{itemize}
			\item we filter the velocity field, using a box filter of size $\ell_1$, and we generate semitransparent surfaces delimitating the domains $\mathcal{D}_1$ where $Q > q_1$;
			\item we filter the velocity field, using a box filter of size $\ell_2 < \ell_1$, and we generate surfaces delimitating the domains $\mathcal{D}_2$ where $Q \geq q_2$, but only if these domains are contained in one of the domains from $\mathcal{D}_1$;
		\end{itemize}
		and this procedure can be used iteratively with several scales (we use at most 3 scales, since the images become too complex for more levels).
			
		Additionally, we wish sometimes to keep track of the relative orientation of the vorticity vectors at the different scales.
		For this purpose we employ a special coloring scheme for the $Q$ isosurfaces:
		for each point of the surface, we compute the cosine of the angle $\alpha$ between the $\ell_2$ filtered vorticity and the $\ell_1$ filtered vorticity:
			\begin{equation}
				\cos \alpha = \frac{(\nabla \times \uu_1) \cdot (\nabla \times \uu_2)}{|\nabla \times \uu_1| |\nabla \times \uu_2|};
			\end{equation}
			the surface is green for $\cos \alpha = 1$, yellow for $\cos \alpha = 0$ and red for $\cos \alpha = -1$, following a continuous gradient between these three for intermediate values.

        The resulting visualisation is available at
        \href{https://www.youtube.com/watch?v=hMMsPGihUi0}{this YouTube link}.

	\section{Observations}

	The opening montage of vortex tubes is very similar to the traditional visualisation: a writhing mess of vortices.
	Upon coarse-graining, additional structure is revealed.
	The large-scale vorticity, which appears as transparent gray, is also arranged in tubes.

	As a next step, we remove all the fine-scale vorticity outside the large-scale tubes.
	The color scheme for the small-scale vorticity is that described earlier, with green representing alignment with the large-scale vorticity and red representing anti-alignment.
	Clearly, most of the small-scale vorticity is aligned with the vorticity of the large-scale tube that contains it.

	We then remove the fine-grained vorticity and pan out to see that the coarse-grained vortex tubes are also intricately tangled and intertwined.
	Introducing a yet larger scale, we repeat the previous operations.
	The relative orientation properties of the vorticity at these two scales is similar to that observed earlier.

	Next we visualize the vortex structures at all three scales simultaneously, one inside the other.
	It is clear that the small vortex tubes are transported by the larger tubes that contain them.
	However, this is not just a passive advection.
	The small-scale vortices are as well being distorted by the large-scale motions.
	
	To focus on this more clearly, we now render just the two smallest scales.
	One can observe the small-scale vortex tubes being both stretched and twisted by the large-scale motions.
	The stretching of small vortex tubes by large ones was suggested by Orszag and Borue \cite{borue_local_1998} as being the basic mechanism of the turbulent energy cascade.
	As the small-scale tubes are stretched out, they are ``spun up'' and gain kinetic energy.
	Here, this phenomenon is clearly revealed.
	The twisting of small-scale vortices by large-scale screw motions has likewise been associated to helicity cascade \cite{eyink_multi-scale_2006}.
	The video thus allows us to view the turbulent cascade in progress.

	Next we consider the corresponding view with three levels of vorticity simultaneously.
	Since the ratio of scales is here 1:15:49 we are observing less than two decades of the turbulent cascade.
	One must imagine the complexity of a very extended inertial range with many scales of motion.

	Not all of the turbulent dynamics is tube within tube.
	In our last scene we visualize in the right half domain all the small-scale vortices, and in the left domain only the small-scale vortices inside the larger scale ones.
	In the right half, the viewer can observe stretching of the small-scale vortex structures taking place externally to the large-scale tubes.
	The spin-up of these vortices must contribute likewise to the turbulent energy cascade.

	\subsection{Acknowledgments}
	This work is supported by the National Science Foundation's CDI-II program, project CMMI-0941530.

%	\bibliography{cited_stuff}
%

\end{document}